\documentclass[12pt]{iopart}

\newcommand{\tip}{$t_{\rm IP}$}
\newcommand{\zip}{$z_{\rm IP}$}
\usepackage{iopams}  
\usepackage{epsfig}
\usepackage{color}
\begin{document}

\hspace{9cm}{  \vbox{
               \hbox{Belle Preprint 2009-1}
\hbox{KEK Preprint   2008-47}
\hbox{arXiv:1001.1194}
\hbox{JINST\_006p\_0110}
}}

\title[KEKB Beam Collision Stability observed with the Belle Detector]{ 
KEKB Beam Collision Stability at the Picosecond Timing and Micron Position
Resolution as observed with the Belle Detector
}


\author{H.\,Kichimi$^1$, 
K.\,Trabelsi$^{1}$, S.\,Uehara$^{1}$,
M.\,Nakao$^{1}$, K.\,Akai$^{1}$, T.\,Ieiri$^{1}$, M.\,Tobiyama$^{1}$, 
M.\,D.\,Jones$^{2}$, M.\,W.\,Peters$^{2}$, G.\,S.\,Varner$^{2}$ and T.\,E.\,Browder$^{2}$
}
\address{$^1$ High Energy Accelerator Research Organization (KEK), Tsukuba, 305-0801 Japan}
\address{$^2$ Department of Physics and Astronomy, University of Hawaii, Honolulu, 96822 USA}
\ead{kichimi@post.kek.jp}

\date{\today}

\begin{abstract} 
Using the Belle detector we study the characteristics of beam collisions at the KEKB
3.5\,GeV\,$e^+$ on 8\,GeV\,$e^-$ asymmetric energy collider.
We investigate the collision timing {\tip} and its $z$-coordinate along the beam axis 
{\zip} as a function of the position of the colliding bunch in a beam train.
The various {\tip} and {\zip}  behaviors observed by Belle are attributed 
to beam loading effects in the radio frequency cavities that accelerate the beams 
with a beam abort gap.
We report these results in detail and discuss the prospects for the SuperKEKB collider.
\end{abstract}

%

\maketitle

\normalsize
\newpage

\normalsize

\vskip 1.0cm

\section{Introduction}\label{sec:intro}
The B-factories, Belle~\cite{belle} at KEKB~\cite{kekb} and BaBar~\cite{babar} at PEPII~\cite{pep2}, 
have carried out precision measurements of various decay modes of $B$ mesons 
to verify the Kobayashi-Maskawa mechanism~\cite{km} that explains CP-violation in the Standard Model.
Belle is a sophisticated $4\pi$ detector, which provides precision 
measurements of the $B$-meson decay vertex, excellent momentum resolution 
and particle identification capability. KEKB is an asymmetric collider 
with beams of 3.5\,GeV\,$e^+$ 
in the low energy ring (LER) and 8\,GeV\,$e^-$ 
in the high energy ring (HER). The BaBar detector at the PEPII collider 
is comparable to the Belle detector at the KEKB collider. 
The total integrated luminosity of the two B-factories is 1.5\,ab$^{-1}$ after ten years of data taking.

KEK plans to upgrade the detector and collider to study CP-violation 
in $B$-decays with extremely high precision to search for new physics beyond the Standard Model,
with an integrated luminosity of at least 50\,ab$^{-1}$~\cite{superbdet}.  
The upgrade plan is to develop an asymmetric $e^+ e^-$ collider with a super high luminosity 
of $8\times10^{35} {\rm cm}^{-2}{\rm s}^{-1}$, and to develop a detector with improved performance 
under higher beam backgrounds (due to beam currents that are a few times larger).
There are significant challenges to developing such a SuperKEKB factory~\cite{superkekb}. 

We report a study of the behaviors of the beam collision timing {\tip} and its position {\zip} along the beam axis
as measured by Belle. These behaviors are finally attributed  to
the beam phases of the colliding bunches in KEKB.
To date, there has been no report on these behaviors, as it requires
a collaboration between experts from the detector and the collider. 
Moreover, the accelerator parameters relevant in this study were measured only during certain periods of
machine study, 
while the Belle measurement is available throughout physics data taking.
In this study, we use 
the time-of-flight detector (TOF), which has a time resolution of 110\,ps, for  the study of the collision timing
{\tip} and the silicon vertex detector (SVD) and the central drift chamber (CDC) for the study of the
collision $z$-coordinate {\zip}, which is measured with a resolution of about 100~$\mu$m~\cite{belle-svd}.
These Belle measurements of {\tip} and {\zip} can be understood in terms of KEKB design parameters
and operating conditions.

Our goal is to understand the sources of systematic uncertainties in the timing measurements with Belle at KEKB,
and to estimate prospects for the Super KEKB-factory, in particular 
for particle identification detectors such as 
the time-of-propagation (TOP)~\cite{superbdet,top} and Cherenkov timing 
detectors~\cite{superbdet,cherenkov}, 
which require precision timing measurements with systematic uncertainty better than 20\,ps.

\subsection{Belle detector}\label{subsec:belle}

\begin{figure}[!htb]
\centering
\includegraphics[width=0.70\textwidth]{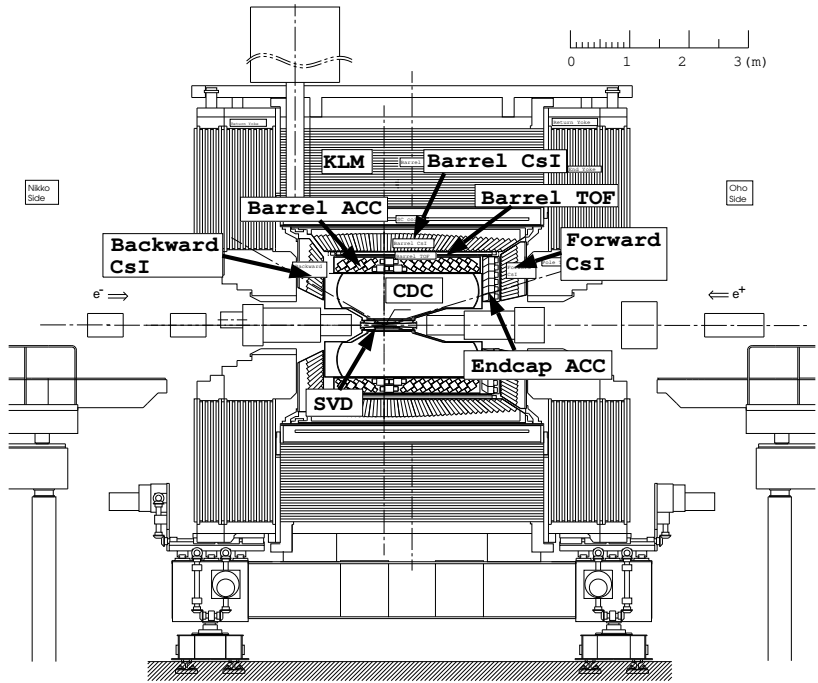}
\centering
\caption{ 
The Belle detector.
}
\label{fig:belle}
\end{figure}

A comprehensive description of the Belle detector is given 
elsewhere~\cite{belle}. 
Figure~\ref{fig:belle} shows the Belle detector configuration.
Charged tracks are reconstructed from hit information 
in a central drift chamber
(CDC) located in a uniform 1.5 T solenoidal magnetic field. The $z$ axis of the
detector and the solenoid are along the positron beam, with the positrons moving
in the negative $z$ direction. The CDC measures the longitudinal and transverse momentum
components (along the $z$ axis and in the $r\phi$ plane, respectively). 
Charged particle trajectory coordinates near the collision point are provided by
the SVD. Photon detection and energy measurements are
performed with a CsI(Tl) electromagnetic calorimeter (ECL). The ECL is also used for
identification of electrons. Identification of pions, kaons and protons is 
performed using information from the time-of-flight counters (TOF) and 
silica-aerogel Cherenkov counters (ACC). The ACC provides good separation 
between kaons and pions or muons at momenta above 1.2~GeV/$c$. The TOF system 
consists of a barrel of 128 plastic scintillation counters and can distinguish
$K/\pi$ for tracks with momenta below 1.2~GeV/$c$. 
Lower energy kaons are also identified using
specific ionization ($dE/dx$) measurements in the CDC. The magnet return yoke
is instrumented to form the $K^0_L$ and muon detector (KLM), which detects muon
tracks and provides trigger signals. 

In order to study the  characteristics of {\tip} and {\zip},
we use a large sample of $\mu$-pair events ($e^+ e^- \to \mu^+ \mu^-$),
which were recorded using the standard Belle trigger 
and selected through the standard Belle data processing.
The accumulation rate of $\mu$-pair events is about 9\,Hz 
at a nominal luminosity of $1.5\times10^{34}{\rm cm}^{-2}{\rm s}^{-1}$.
The purity of the selected $\mu$-pair events is better than 98\%.

\subsection{KEKB}\label{subsec:kekb}
The KEKB collider achieved a peak luminosity of $2.1\times10^{34}{\rm cm}^{-2}$s$^{-1}$ in June 2009, 
which is more than twice the design value of $1\times10^{34}{\rm cm}^{-2}$s$^{-1}$.
An integrated luminosity of 1018\,fb$^{-1}$ was accumulated with the Belle detector by December 2009. 

\begin{figure}[!htb]
\centering
\includegraphics[width=0.70\textwidth]{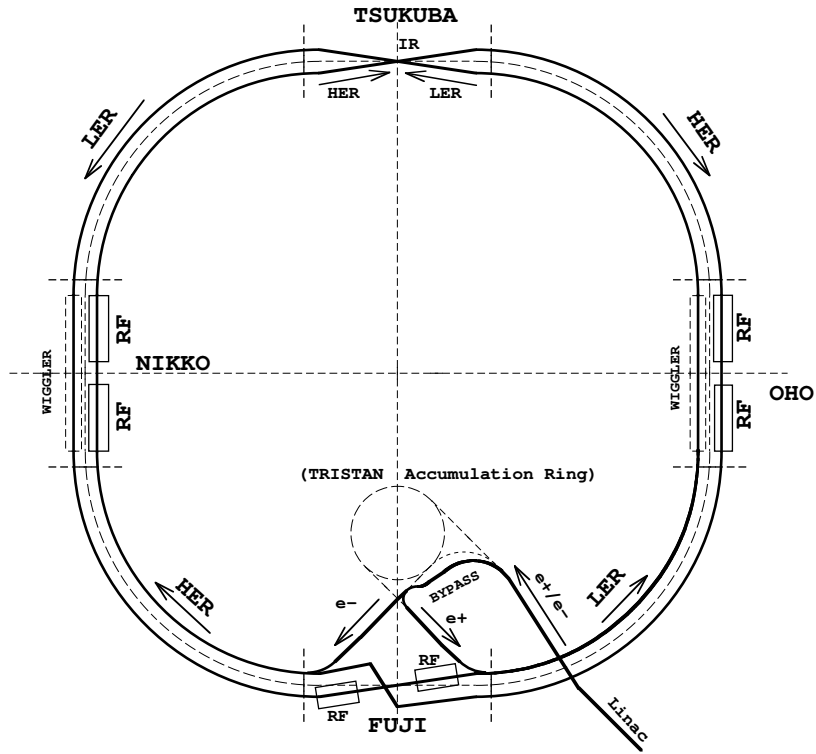}
\centering
\caption{ 
The KEKB collider complex.
}
\label{fig:kekb}
\end{figure}

Figure~\ref{fig:kekb} shows the KEKB collider complex~\cite{kekb}, which has a circumference of approximately 3\,km.
KEKB stores beam currents up to 2.0\,A in the LER and 1.35\,A in the HER
with 508.9\,MHz radio frequency (RF) acceleration systems, which are located at the FUJI, OHO and NIKKO straight sections.
The RF systems consist of two types of 
higher order modes damped cavities: the normal-conducting Accelerator Resonantly-coupled with 
Energy-Storage (ARES) cavities, and Superconducting (SC) cavities. Twenty ARES cavities are used 
in the LER to handle higher beam currents, while twelve ARES and eight SC cavities are used 
in the HER to achieve a higher voltage~\cite{kekb1}.
The RF reference signal, generated at the KEKB control room near the FUJI hall, is distributed 
through two coaxial cable lines in the clock- and anti-clock-wise directions around the KEKB collider.
{These cables have good phase versus temperature characteristics, less than 10 ppm/$^{\circ}$C. The RF reference signals are
phase-locked at each station and distributed to the RF cavities. Thus,
the accuracy of the 508.9\,MHz RF reference clock is found to be typically 6\,ps in one circulation 
around the KEKB collider.} The Belle detector 
uses the same RF reference signal distributed at TSUKUBA hall.

\begin{figure}[!htb]
\centering
\includegraphics[width=1.0\textwidth]{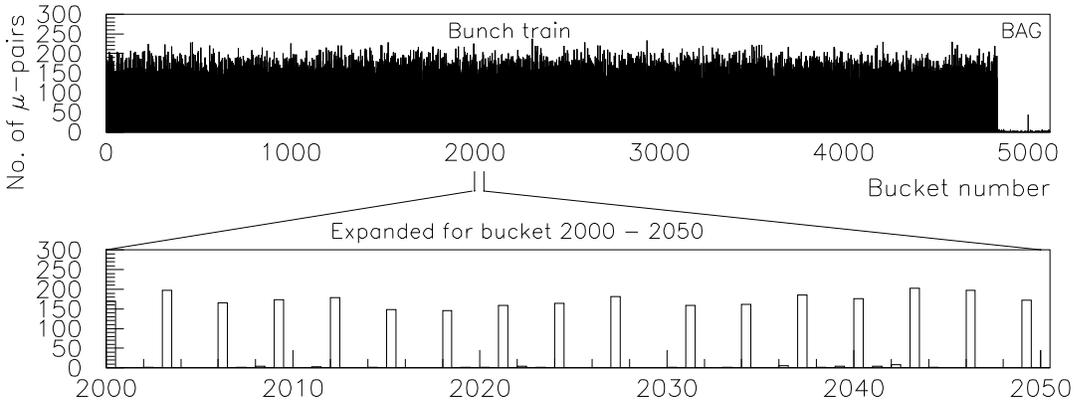}
\centering
\caption{ 
The number of $\mu$-pair events as a function of the RF bucket number,
which are reconstructed by the Belle detector for Exp61. 
{See Appendix~\ref{subsec:a1} for details.}
}
\label{fig:exp61bucket}
\end{figure}

The LER and HER beams collide bunch-by-bunch at the IP with a finite angle of $22$~mrad in the horizontal plane. 
Each beam train consists of 5120 buckets synchronous to the 508.9\,MHz RF clock.
{In usual operation, the buckets are filled with three- or four-bucket spacing and with a period of 49 RF buckets.
A typical beam train contains 1584 bunches (buckets filled by beam).} 
The beam train is followed by a beam abort gap (BAG) of 270 empty buckets,
which is needed to safely kick out the beam from the ring.
Figure~\ref{fig:exp61bucket} shows a typical collision bunch pattern
as a function of bucket number,
which is obtained from analysis of $\mu$-pair events accumulated with Belle.
Here the number of $\mu$-pair events is shown as a function of the bucket number.
We clearly see the bunch train followed by the BAG as described above.

\begin{figure}[!htb]
\centering
\includegraphics[width=1.0\textwidth]{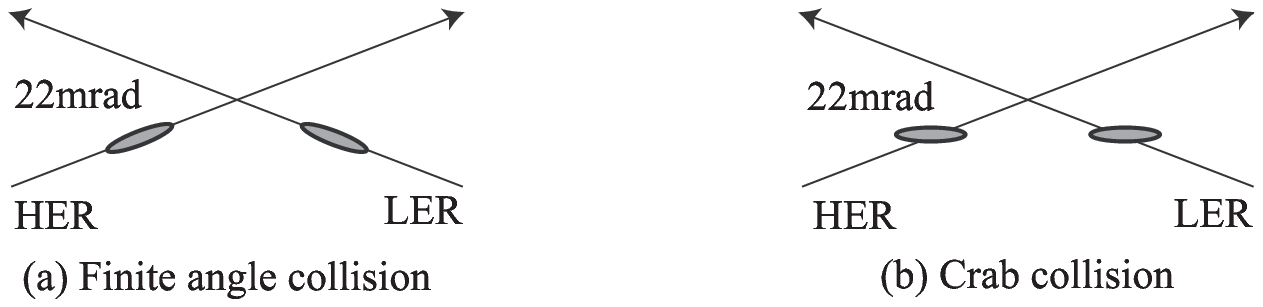}
\centering
\caption{ 
{
Collision schemes at KEKB: (a) finite angle collisions and (b) crab crossing.
}}
\label{fig:collisionschemes}
\end{figure}

In December 2006, two superconducting crab cavities~\cite{kekb2} were installed in the LER and the HER, one for each, 
in order to operate KEKB in a crab crossing scheme aiming to increase luminosity.
{Figure~\ref{fig:collisionschemes} illustrates (a) finite angle collisions and (b) crab crossing at the IP.}
In the crab crossing scheme the bunches are kicked in the horizontal plane by transverse RF in the crab cavities
{(located in the NIKKO straight sections)} 
so that they rotate and collide head-on at the IP (TSUKUBA) even in finite-angle crossing beam orbits.


\subsection{Belle data sets used in this analysis}\label{subsec:data}
We study three data sets that are summarized in 
Table~\ref{run-summary} by the run period, LER and HER beam currents, 
recorded peak luminosity, CM energy and status of crab cavities. 
(1) Exp55 is the data set before crab cavity installation (no-crab crossing), in which
KEKB achieved the highest peak luminosity with the highest LER and HER currents.
(2) Exp61 is the data after the crab cavity installation (crab crossing), during which
KEKB was operated with head-on collisions. 
(3) Exp65 is the Spring 2008 run period, in which we recorded data at the
center-of-mass (CM) energies of the $\Upsilon$(4S) and $\Upsilon$(1S) resonances. 
{To change the CM energy, both the LER and HER beam energies are changed proportionally.}
In Table~\ref{run-summary}, $\Upsilon$(1S),   $\Upsilon$(4S) and $\Upsilon$(5S)
indicate the CM energies of 9.460\,GeV, 10.580\,GeV, 10.860\,GeV, respectively.
Here $\Upsilon$(6S) stands for the energy region just above the $\Upsilon$(5S).
We use more than 300\,k $\mu$-tracks for each data sample.

\begin{table}[!htb]
\caption{ Summary of the three run periods (Exp55, Exp61 and Exp65) used in this analysis. 
The peak luminosity is in units of 10$^{34}$ cm$^{-2}$s$^{-1}$.}
\centering
\begin{tabular}{|c|c|c|c|c|c|c|} \hline
    & Run period  & LER & HER & Peak lum.  & CM energy & Crab \\ \hline
Exp55  & Sept.-Dec. 2006 & 1.65\,A & 1.37\,A  &  1.7 & $\Upsilon$(4S) & - \\ \hline 
Exp61  & Oct.-Dec. 2007  & 1.62\,A & 0.81\,A  &  1.4 & $\Upsilon$(4S,5S,6S) & ON \\ \hline 
Exp65  & Apr.-June 2008  & 1.61\,A & 0.95\,A  &  1.6 &  $\Upsilon$(4S,1S)   & ON \\ \hline 
\end{tabular}
\label{run-summary}
\end{table}

\section{Collision timing and collision position}\label{sec:iptipz}
\subsection{Variation of Collision timing {\tip}}\label{sec:ipt}
We first discuss the long term behavior of the collision timing {\tip}.
We define {\tip}=$(t_{\rm obs}-t_{\rm exp})$ using $\mu$-pair events.
Here, $t_{\rm obs}$ is the measured time-of-flight for one $\mu$ candidate
and $t_{\rm exp}$ is the prediction from its track trajectory in the CDC
(see Appendices~\ref{subsec:a1} and \ref{subsec:a2}).
We usually obtain a mean {\tip}=$t_{\rm obs}-t_{\rm exp}=0$ for each 
run using the standard TOF calibration after run-by-run correction. 
For online data processing, we use a default database obtained by TOF calibration
using $\mu$-pair events from the previous run period.
Once the database is installed at the beginning of the new run period,  
it is unchanged during the whole run period, typically for several months.
{If both the KEKB collider and the Belle detector (TOF and CDC) are stable, 
the mean {\tip} value should be unchanged.  
We conclude that the variation of the online {\tip} value must be due to changes 
in beam conditions or
changes in the RF reference timing from KEKB to the TOF readout system. }

\begin{figure}[!htb]
\centering
\includegraphics[width=1.0\textwidth]{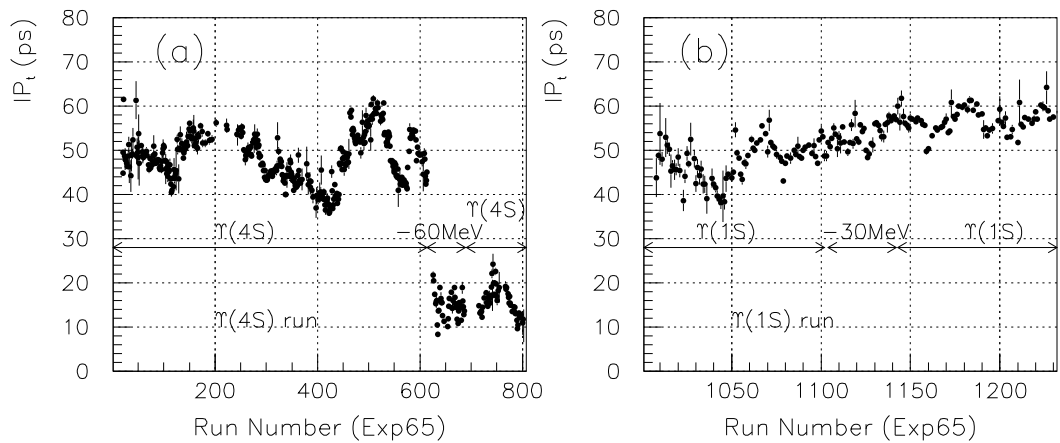}
\centering
\caption{ {\tip} versus run number for Exp65. 
(a) $\Upsilon$(4S) runs and (b) $\Upsilon$(1S) runs.  
The run number is reset to 1001 for $\Upsilon$(1S) runs
after run 810.
}
\label{fig:dt-vs-run}
\end{figure}

Figure~\ref{fig:dt-vs-run} shows the mean {\tip} as a function of run number for Exp65, 
taken at the energies of the (a) $\Upsilon$(4S) and (b) $\Upsilon$(1S) from April to June 2008.
We see a slow drift within a $\pm10$\,ps range and step-like shifts at runs 610 and 810. 
The {\tip} changes are due to various effects in the KEKB collider and the Belle detectors.
While the individual sources are not understood completely, the {\tip} variation can be corrected
even with significant changes of beam energy and frequent optics corrections 
during KEKB operation. In practice, we apply a run-by-run correction for {\tip} to 
reduce the bias for physics analysis.
We have a time resolution of 110\,ps for the TOF system (including a readout system error of 40\,ps)
after ten years of operation, starting with a value of 96\,ps in 1999  (see Appendix~\ref{subsec:a2}). 

In the analysis that follows, we use the {\tip} value 
averaged over a run period and corrected for the run-by-run variation. Here,
a run typically extends for a few hours to a day and contains about 300\,k $\mu$-pair events.
We normally assume that the collision timing is independent of the bunch position in a beam train.  
However, we observe a fine structure in the collision timing as discussed in the following sections.

\subsection{{\tip} and {\zip} vs bucket number by Belle}\label{subsec:iptipz-bucket}

The position of each beam bunch may be displaced from
that synchronized with the RF frequency, because of
dynamic  beam loading effects in the RF cavities and the longitudinal impedance of other accelerating components.  
Therefore, it is interesting to study {\tip} and {\zip} as a function of the bucket number  $N_{\rm bucket}$,
the collision bunch position in a bunch train.

\begin{figure}[!htb]
\centering
\includegraphics[width=1.0\textwidth]{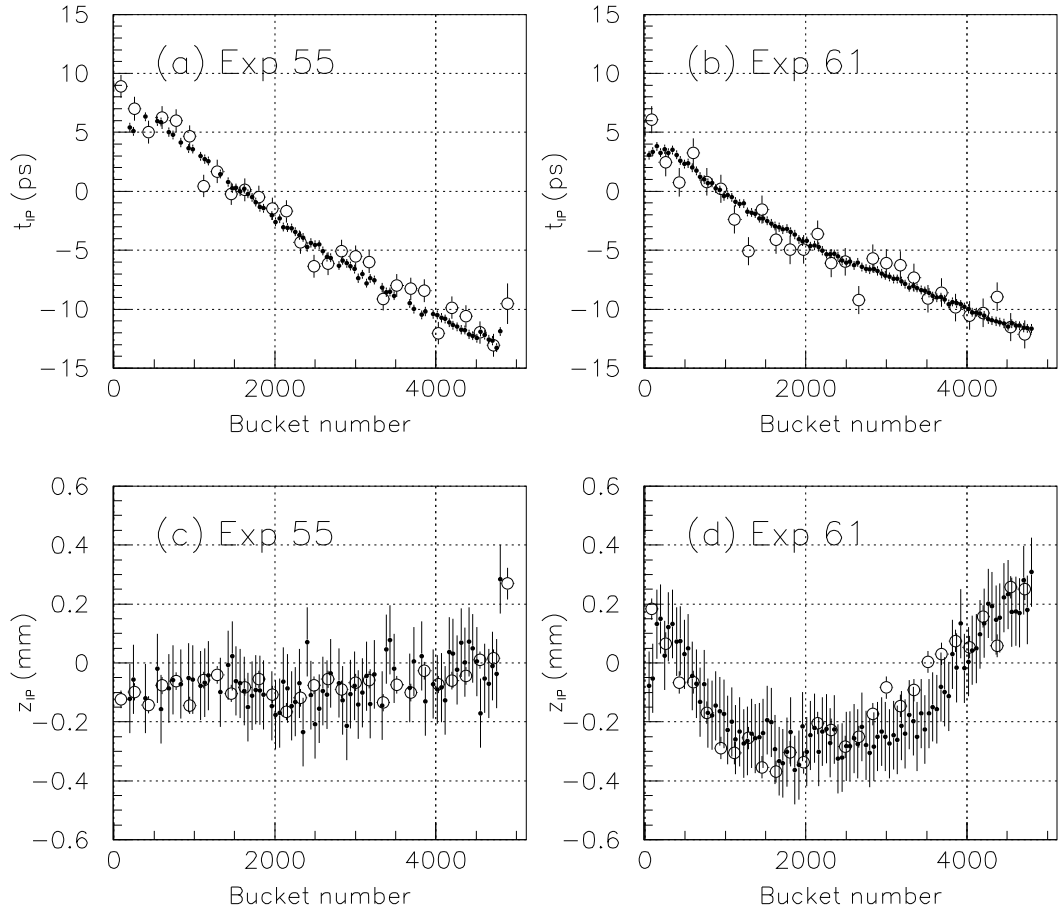}
\centering
\caption{
{\tip} and {\zip} vs bucket number; (a) and (c) for Exp55, and (b) and (d) for Exp61, respectively.
The open circles are Belle measurements, while the filled circles are
predictions from the measured LER and HER beam phases (see text).
}
\label{fig:ipt-ipz-exp5561}
\end{figure}

Figure~\ref{fig:ipt-ipz-exp5561} shows the Belle measurements of {\tip} and {\zip} as a function of
$N_{\rm bucket}$; (a) and (c) show the results for Exp55
while (b) and (d)  are the corresponding results for Exp61.
The open circles indicate the Belle measurements.
We see linear {\tip} variation both for Exp55 and Exp61.
On the other hand, we observe parabolic {\zip} behavior for Exp61, while {\zip} is almost flat 
for Exp55.  These Belle observations of {\tip} and {\zip} behaviors are discussed
quantitatively in the next section.  

\subsection{Beam phase and transient beam loading at KEKB}\label{subsec:beamphase-kekb}

\begin{figure}[!htb]
\centering
\includegraphics[width=1.0\textwidth]{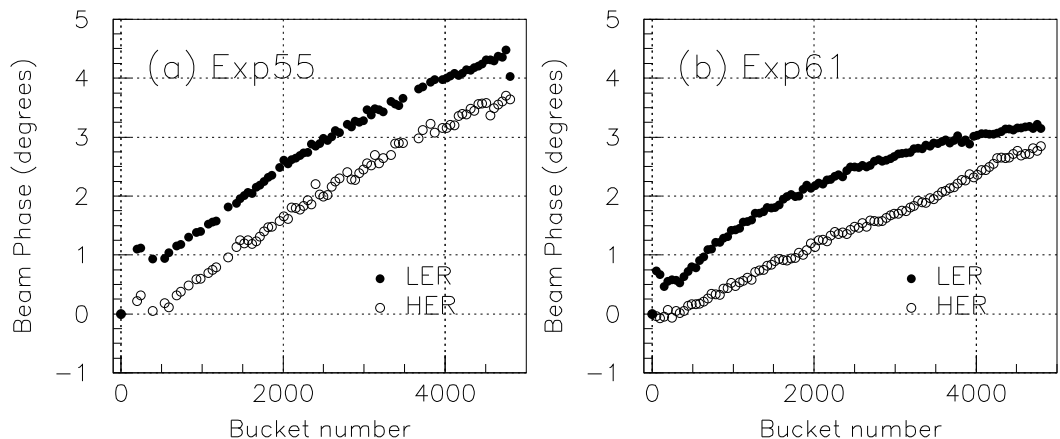}
\centering
\caption{
Beam phase as a function of bucket number for 
(a) Exp55 and (b) Exp61.
}
\label{fig:kekb-beam-phase}
\end{figure}

Figure~\ref{fig:kekb-beam-phase} shows the beam phases 
as a function of the bucket number (N$_{\rm bucket}$) measured by KEKB
in the same periods of (a) Exp55 and (b) Exp61. 
The beam phase is defined as the phase with respect to the corresponding RF bucket.
Here, the phases are measured at every 49-th bucket by the gated beam position monitor 
(GBPM)~\cite{bpmkekb} (see Appendix~\ref{subsec:b1}), and
the phase of the first bunch after the beam abort gap (bucket number 0) is set to zero by definition.
We observe a beam phase advance with increasing bucket number in the LER and the HER 
with a maximum phase change of approximately 3.5 degrees.
Figure~\ref{fig:kekb-beam-phase}~(a) shows a similar pattern of beam phases in the LER and the HER 
for Exp55, while
Fig.~\ref{fig:kekb-beam-phase}~(b) shows a systematic difference between them for Exp61.  

The observed beam phase modulation is known as "transient beam loading" in the
RF cavities due to the existence of
a beam abort gap (BAG) after a bunch train. The beam phase change between the head and tail bunches is
approximately proportional to the total beam current and the length of the BAG.
The modulation pattern is sensitive to the type of RF cavities (ARES or SC) and the total beam current,
as well as the pattern of bunch currents in the bunch train (see Section~\ref{sec:disc} and Appendix~\ref{subsec:b2}).

\subsection{Effect of beam phases to {\tip} and {\zip}}\label{subsec:prediction}

In this section, we give a quantitative discussion of the {\tip} and {\zip} behaviors observed by Belle
{from beam phases of the collision bunches in the beam train.} 
Assuming head-on collisions of the LER and HER bunches in the horizontal plane at the interaction point (IP)
for a finite-angle collision with crab crossing, we derive the following formulae,
\begin{equation}
t_{\rm IP}={T}_{\rm RF}\times\frac{-(\phi_{\rm HER}+\phi_{\rm LER})}{2}\times\frac{1}{360}~~{\rm (ns)}  \label{eq1}\\
\end{equation}
\begin{equation}
z_{\rm IP}={L}_{\rm RF}\times\frac{(\phi_{\rm HER}-\phi_{\rm LER})}{2}\times\frac{1}{360}~~{\rm (m)}   \label{eq2}
\end{equation}
where ${T}_{\rm RF}$ (approximately 2\,ns) is one RF period and 
${L}_{\rm RF}$ (approximately 0.6\,m) is one bucket interval given by $c{T}_{\rm RF}$ where $c$ is the speed of light.
The beam phases $\phi_{\rm HER}$ and $\phi_{\rm LER}$ are in degrees.
{
Here we assume that the change of {\zip} is much smaller than the  
vertical beta function $\beta^*_y$ of 6\,mm at the IP. If this is not the  
case, the change of {\zip} will be smaller than the value given by Eqn.~(\ref{eq2}) due to the hourglass effect.}
The collision timing of the tail bunch is advanced by about 20\,ps relative to the synchronous RF
bucket, so that {\tip} is delayed by the same amount.
It should be noted that the {\zip} behavior is very sensitive to the LER and HER beam phase difference,
while the {\tip} variation is almost proportional to the beam phase modulation.

In finite-angle collisions without crab cavities, 
the bunches are more separated in the horizontal plane before and after the collision than in the case of the head-on collisions with crab cavities,
{as seen in Fig.~\ref{fig:collisionschemes}.
We estimate the {\zip} variation by a naive calculation
from the geometrical overlap of the colliding LER and HER bunches in three dimensions as a function of time,
and taking into account finite-angle collisions with and without crab cavities and the beam focusing optics at the IP.
We find that the {\zip} variation is slightly smaller, about 70\% of that predicted by Eqn.~(\ref{eq2}). }

Figure~\ref{fig:ipt-ipz-exp5561} shows the Belle measurements (open circles) of {\tip} and {\zip} 
compared with the predictions (filled circles) from Eqns.~(\ref{eq1}) and (\ref{eq2}).
{They are in good agreement 
except for a small systematic difference in Fig.~\ref{fig:ipt-ipz-exp5561}(d).}
In Exp55, the LER and HER beam phase variations are similar, 
so that the {\zip} behavior is flat and has no distinctive structure.
In the case of Exp61, we see a parabolic {\zip} behavior with a range of approximately 0.6\,mm,
due to the significant difference between the LER and HER beam phases. 
{This is attributed to the difference in the LER beam phase behaviors between Exp55 and Exp61
as shown in Fig.~\ref{fig:kekb-beam-phase}.
The cause of the difference is discussed in Section~\ref{sec:disc}.}

Thus, the observed {\tip} and {\zip} behaviors are explained well 
by the beam phase change as a function of the bunch position in the colliding bunch trains (beam phase modulation), 
which is caused by the transient beam loading effect in the RF system due to the beam abort gap in the beam train.

\subsection{Size of Collision point {\zip}}\label{subsec:beamsize}

\begin{figure}[!htb]
\centering
\includegraphics[width=1.0\textwidth]{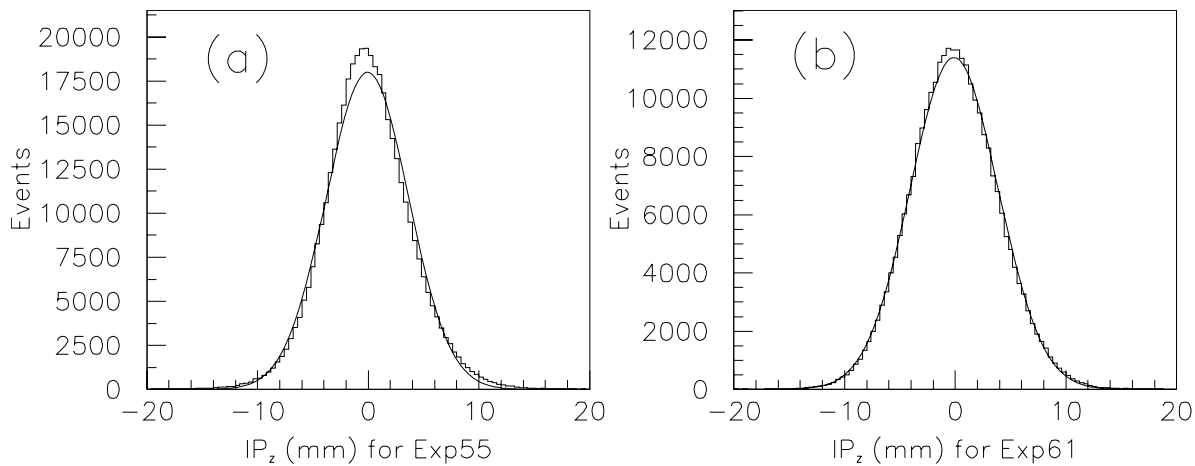}
\centering
\caption{ 
{\zip} distributions for (a) Exp55 and (b) Exp61.
The Gaussian widths are 3.71\,mm and 3.96\,mm, respectively. 
}
\label{fig:ipz-e55-e61}
\end{figure}

Figures~\ref{fig:ipz-e55-e61} (a) and (b) show the {\zip} distributions
for Exp55 and Exp61, respectively.
The widths of these distributions, fitted to a Gaussian function with 
$\sigma=3.71$\,mm and 3.96\,mm, respectively, are affected by 
the LER and HER bunch sizes along the beam axis and the collision scheme (no-crab
or crab crossing). 

The natural bunch lengths for (LER, HER) are estimated to be (4.86\,mm, 4.84\,mm) 
for Exp55 and (4.86\,mm, 5.22\,mm) for Exp61 from the known KEKB operation parameters (Table~\ref{ipz-spread}).
The last two columns give the expected {\zip} widths and the corresponding Belle measurements.
Here, the {\zip} width is estimated from the natural bunch lengths by 
$\sigma_{\rm IP}=1/\sqrt{1/\sigma_{\rm LER}^2+1/\sigma_{\rm HER}^2}$.  
{The widths measured by Belle are about 10\,\% larger than these  
estimates, due to the bunch lengthening effect caused by  
the inductive part of the longitudinal impedance of the ring components.}
The {\zip} widths in the crab crossing runs (Exp61) seem to be larger than those in the no-crab crossing runs (Exp55). 

\begin{table}[!htb]
\caption{ Size of collision position {\zip} and LER and HER bunch sizes}
\centering
\begin{tabular}{|c|c|c|c|c|c|} \hline
&\multicolumn{2}{c|}{natural bunch length} & {\zip} width & {\zip} width & crab \\ \cline{2-5}
    & LER   & HER   &  KEKB est. & Belle meas. & status \\ \hline  
Exp55  & 4.86 mm & 4.84 mm & 3.43 mm & 3.71 mm & OFF \\ \hline 
Exp61  & 4.86 mm & 5.22 mm & 3.56 mm & 3.96 mm & ON  \\ \hline 
\end{tabular}
\label{ipz-spread}
\end{table}

\section{Discussion on the LER beam phase behaviors}\label{sec:disc}

{
Here, we discuss the cause of the significant difference in the LER beam phase behaviors in Exp55 and Exp61
due to beam loading in the RF cavities (see Appendix~\ref{subsec:b2}).
We first consider the possibility that the striking difference is related to the crab cavity 
installation.
However, as crab cavities of the same type were installed in both the LER and HER,
we cannot explain why we observed a significant difference only in the LER beam
as shown in Fig.~\ref{fig:kekb-beam-phase}.}

{In general,} the beam phase change with bunch position is attributed to transient beam loading 
in the RF cavities, which is sensitive to the total beam current and the bunch current pattern after the BAG. 
The transient beam loading effect is characterized by the ratio ${T}_0/{T}_f$, 
where ${T}_0$ is the beam revolution period and ${T}_f$ is the filling time of the RF cavity. 
In the limit ${T}_0\gg{T}_f$, the beam phase shift (relative to the RF bucket) is expected to be largest in the beam head 
just after the BAG.
This results in non-linear behavior of the beam phase.
On the other hand, in the limit ${T}_0\ll{T}_f$, the beam phase is expected to be linear with 
bunch position as the effect is averaged over many beam revolutions.
ARES cavities have a ${T}_0/{T}_f$ value of 0.54, which is about twice as large as SC cavities.
Therefore, the LER, which consists of ARES cavities only, tends to have stronger non-linear behavior of the beam phase.
\begin{figure}[!htb]
\centering
\includegraphics[width=1.0\textwidth]{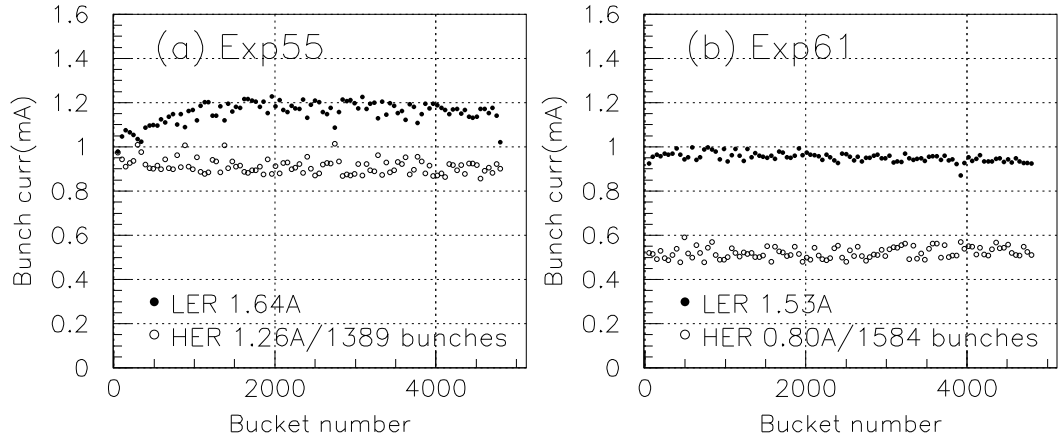}
\centering
\caption{
Bunch current at each 49-th bucket for 
(a) Exp55 and (b) Exp61.
}
\label{fig:ibunch}
\end{figure}

The above discussion is based on the assumption that the bunch current (current intensity) is uniform along the beam train.
Figure~\ref{fig:ibunch} shows the LER and HER bunch current variation versus bunch position for
(a) Exp55 and (b) Exp61.
We see that the HER bunch currents are constant in both; the HER beam phases vary 
almost linearly with bunch position as shown in Fig.~\ref{fig:kekb-beam-phase}.
Using Eqn. (\ref{eq4}), we calculate the ratio of the head to tail phase shifts to be 0.74 for Exp61/Exp55 from 
an HER current of 0.80\,A (1.26\,A) and an RF voltage of 13\,MV (15\,MV).
We also measure an HER beam phase shift of 2.8 (Exp61) and 3.7 (Exp55) degrees, and find
the ratio of 0.76, which is in good agreement with the above estimate.

On the other hand, the LER bunch current in Exp55 shows a slow increase of about 20\% 
in the head part (up to bucket number 1000), while it is almost uniform in Exp61. 
We conclude that the non-linear behavior of the beam phase (Exp61) can be attributed to 
the transient response of the ARES cavities to the uniform bunch currents along the bunch train. 
In the case of Exp55, the pattern of non-uniform bunch currents weakens the steep increase 
of the beam phase in the head part and increases the phase change in the tail part, 
and thus gives almost linear phase behavior.

The SuperKEKB~\cite{superkekb} collider was initially designed to store currents of 9.4\,A in the LER and 4.1\,A in the HER (High Current scheme), 
which are four times larger than those of KEKB. 
To reduce the beam-loading effect due to the beam abort gap, which is proportional to the beam currents, 
we have designed a gap length of 2\% (5\% at KEKB) that will require development of new beam abort kickers.
Recently an alternative design called
the Nano-Beam scheme has been proposed and is the baseline option.
This scheme obtains high luminosity via small beam sizes, which are achieved by reducing emittances and $\beta$-functions. 
The required beam currents of 3.3 A (LER) and 1.9 A (HER) are smaller than in the initial SuperKEKB proposal and hence 
the beam loading effects are somewhat reduced.

\section{Summary}\label{sec:summary}

Using the Belle detector, we study the characteristics of colliding beams at the KEKB asymmetric energy collider.
The behaviors of the collision timing ({\tip}) and its $z$-coordinate along the beam axis ({\zip})
observed by Belle are well explained by the beam phase variations along the LER and HER beam trains
(beam phase modulation). Here, we calibrate for the long term variations of {\tip}
(see subsection~\ref{sec:ipt}) whose causes are not well understood. 
The studied beam phase variations are caused by transient beam loading in the RF systems due to 
the beam abort gap in the beam train. The {\zip} behavior is very sensitive to the beam phase difference between LER and HER
beam trains. The beam phase variation is sensitive to the bunch current pattern along the beam train as well.
In collisions of LER 1.6\,A and HER 1.2\,A beams with a 5\,\% beam abort gap at KEKB,
we find an {\zip} variation of about 0.6\,mm, and an {\tip} variation of about 20\,ps.   
To reduce the beam loading effect,
we have designed a gap length of 2\% to allow the SuperKEKB collider 
to store currents of 9.4\,A (LER) and 4.1\,A (HER) for the High Beam current scheme. 
The beam loading effect is much reduced with the same gap design 
for the Nano Beam scheme.

We are able to reduce the systematic errors in collision timing and position due to the beam loading in the RF system
to insignificant levels by correcting the data for the fine structure as a function of the bucket number,
as discussed in this report. 
A crucial issue for the SuperKEKB factory detector, which must realize a systematic error of less than 20\,ps,
will be to reduce time jitter in the readout system synchronized with the RF clock 
and in the calculation of the time-of-flight value from tracking information.

\vspace {0.5cm}
\ack
We are grateful to the Belle collaboration and the KEKB members 
who made this study possible and fruitful.

\appendix
\section{Event timing determination with the TOF detector}\label{sec:a}
\subsection{Bucket number determination}\label{subsec:a1}

A detailed description of the TOF readout system and analysis procedure 
is found in Ref.~\cite{belle-tof}. Here, we explain the TOF timing measurement scheme with
the Time Stretcher~\cite{rd100} (TS) and TDC (TS-TDC), as
it is essential to determining the bucket number of a colliding bunch for each event. 

\begin{figure}[!htb]
\centering
\includegraphics[width=0.8\textwidth]{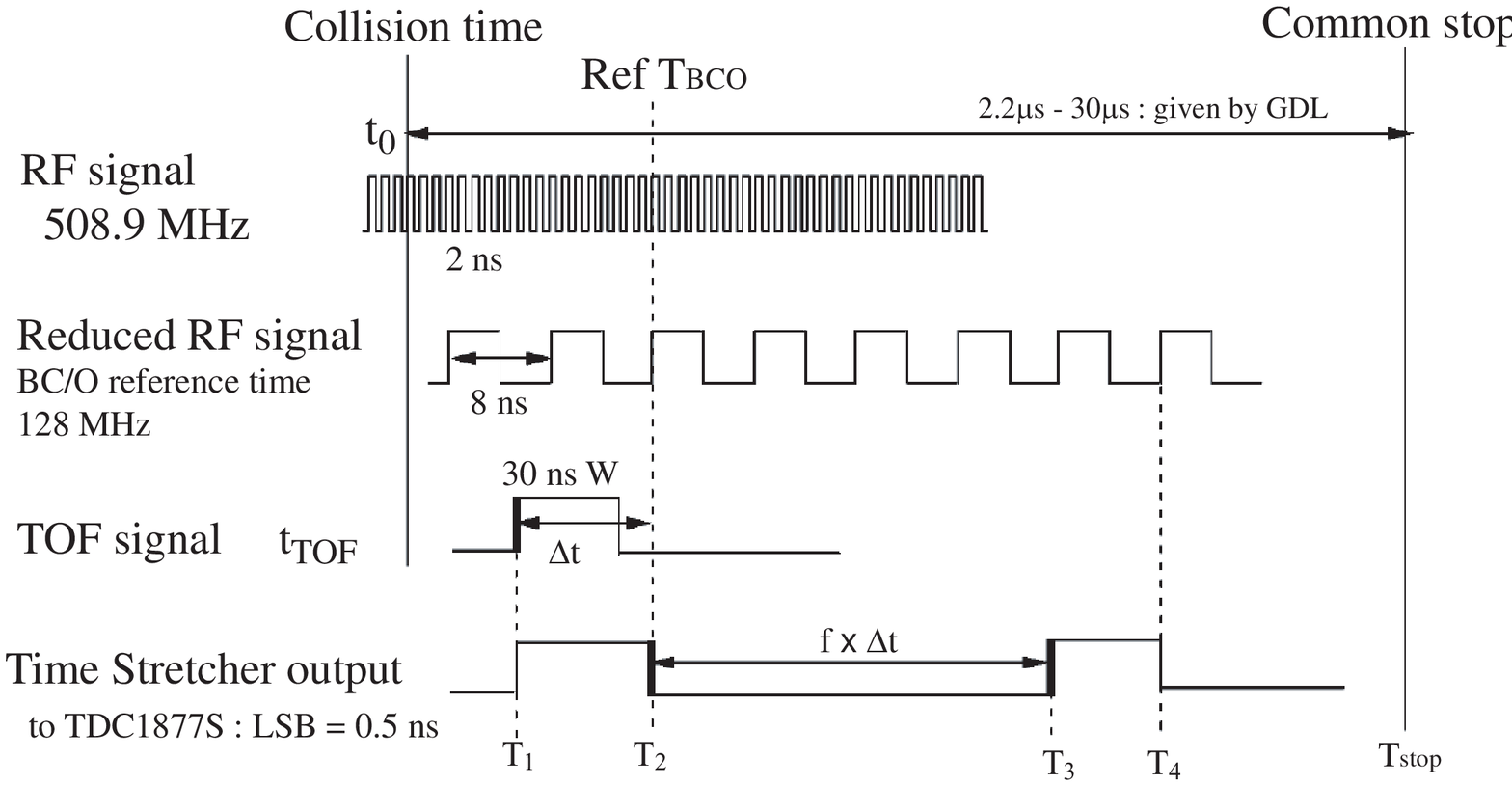}
\centering
\caption{ 
Time Stretcher TDC scheme for TOF. The TS reference clock of 
approximately 8 ns period is generated from the KEKB RF signal of 508.9 MHz. 
}
\label{TS}
\end{figure}

The TOF readout system records a set of charges $Q_i$ and timings $T_i$ from
the rising edges of discriminator outputs for each PMT signal from the TOF detector.
Figure~\ref{TS} shows the block diagram of the TS-TDC scheme. The TS circuit finds the first rising edge $T_2$ of the TS 
reference (reduced RF) clocks following the rising edge $T_1$ of the TOF signal, and expands the time interval $T_{21}=T_2-T_1$
by a factor of approximately $f=20$ to obtain $T_3$ ($T_3-T_2$=$T_{21}\times f$+$a$).  
These times are then read out with the Belle standard TDC with a 0.5~ns least significant bit (LSB), 
providing a 25\,ps LSB resolution. 
We reconstruct the TOF timing $T_{\rm TS-TWC}$ from timings, $T_1$, $T_2$ and $T_3$, using the parameters $f$ and $a$ calibrated 
for each TS channel.
We further apply a Time Walk Correction (TWC) to correct for timing variation proportional to $1/\sqrt{Q}$~\cite{belle-tof}.

The TS reference clock of approximately 8\,ns (127.2\,MHz) is generated from the RF clock of 508.9\,MHz.
As the RF clock is used for the entire KEKB timing control of the LER and HER beams, 
the TS clock is necessarily synchronized with the beam collision timing. 

Each event is triggered with a DAQ clock of approximately 16\,ns period
synchronized with every 8-th RF bucket.
An RF bucket number $n_{\rm DAQ}$ (a multiple integer of 8) corresponding to the trigger is then recorded.
In the standard TOF offline analysis, we determine a precise time relative to the DAQ clock.
In particular, we determine an integer $n_{\rm TOF}$ (a fine bucket number relative to $n_{\rm DAQ}$),
by taking into account track momenta and their path lengths to the TOF counter (times-of-flight with 
$e, \mu, \pi$, kaon and proton mass hypotheses).
We obtain $n_{\rm TOF}$ with a candidate set of the mass assignments for each track,
by optimizing a sum of squared time differences 
$\sum_{ij}{(\delta T_{ij})}^2$ for all tracks in the event, 
where $\delta T_{ij}=(T_{{\rm TS-TWC}{ij}}-n_{\rm TOF}\times T_{\rm RF})-t_{{\rm exp}{ij}}$
is the time difference using the $j$-th mass hypothesis for the $i$-th track.
Here, $T_{\rm RF}$ is the RF period of approximately 2~ns.
Thus, we find $n_{\rm TOF}$, and obtain the measured TOF timing
$t_{{\rm obs}i}=(T_{{\rm TS-TWC}i}-n_{\rm TOF}\times T_{\rm RF})$ for the track
and the bucket number $N_{\rm bucket}=n_{\rm DAQ}+n_{\rm TOF}$ of the collision bunch in the train.
Figure~\ref{fig:exp61bucket} shows the number of 
$\mu$-pair events as a function of the reconstructed bucket number.

\subsection{Time resolution of the TOF detector}\label{subsec:a2}

\begin{figure}[!htb]
\centering
\includegraphics[width=1.0\textwidth]{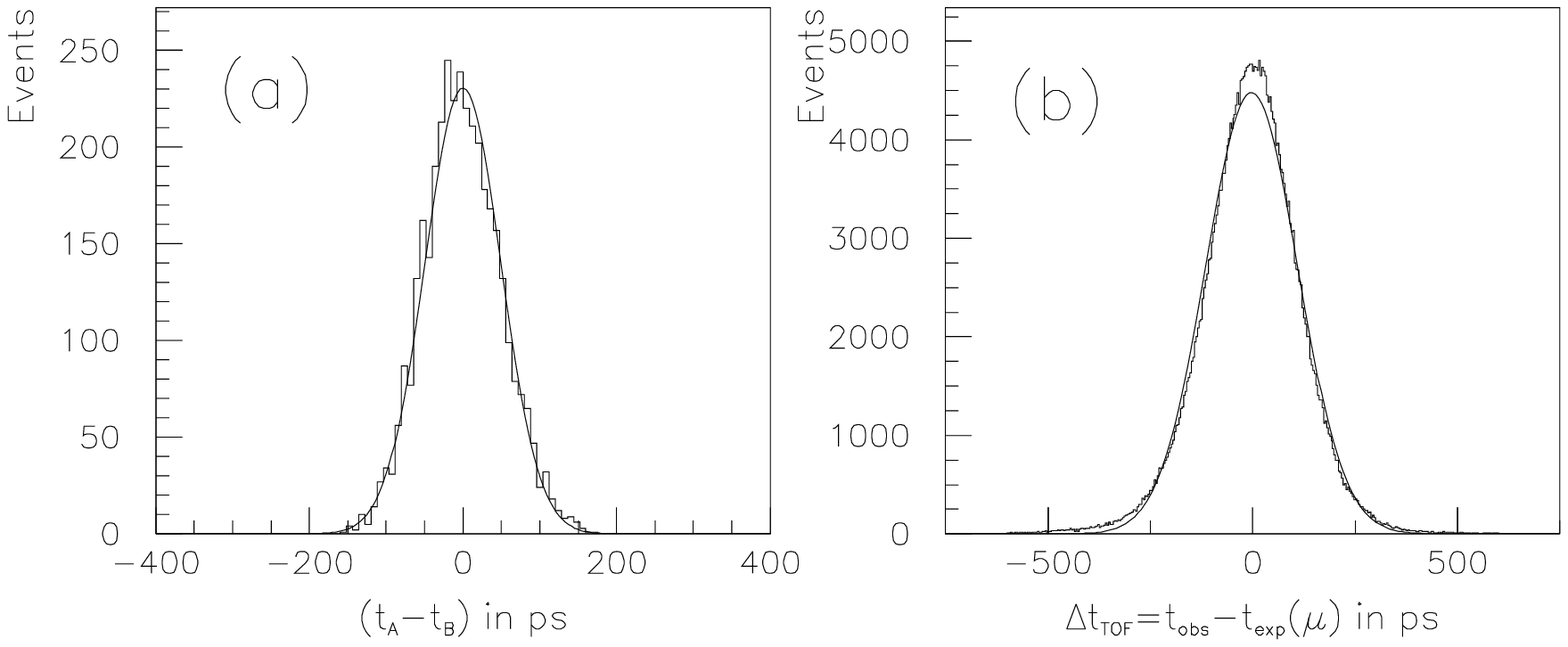}
\centering
\caption{ (a) Distribution of $(t_{\rm A}-t_{\rm B})$ for times measured by 
two independent readout channels A and B for the same input signal. 
(b) Distribution of ${\Delta t}_{\rm TOF}=t_{\rm obs}-t_{\rm exp}$ for $\mu$-pair events for Exp 65,
which gives a TOF time resolution of $\sigma$= 110~ps.
}
\label{fig:time-jitter}
\end{figure}

Here, we discuss the sources of the time resolution for the TOF detector.
Figure~\ref{fig:time-jitter}~(a) shows the distribution of time difference $(t_{\rm A}-t_{\rm B})$
between two measurements with independent channels $t_A$ and $t_B$
for the same input signal, in order to study the time jitter in the TOF readout system.
We estimate a time jitter of 35\,ps from $\sigma(t_{\rm A}-t_{\rm B})/\sqrt{2}$,
which includes the time jitter in the discriminators of the TOF signal,
errors in the TS (20~ps), and errors in the FASTBUS-TDC system. 
Figure~\ref{fig:time-jitter}~(b) shows the distribution of time difference 
$\Delta t=(t_{\rm obs}-t_{\rm exp})$ (={\tip}) for $\mu$-pair events.
Here, $t_{\rm obs}$ is the measured time with TWC calibration after TS reconstruction,
while $t_{\rm exp}$ is the predicted time calculated from the track trajectory in the CDC and 
its extrapolation to the TOF. 

Whenever both KEKB operation and the Belle detector (TOF and CDC)
are stable, the mean of $\Delta{t}$ should be zero.
However, $\Delta{t}$ varies run-by-run and in the long term. 
In practice, we correct for the run-by-run variation to set
the mean $\Delta{t}$ to zero for each run,
so that we have no bias in physics analysis.

The present time resolution of the TOF system is found to be 110~ps for $\mu$-pair tracks.
This resolution includes a systematic error of 40\,ps consisting of the following components; 
a time jitter of 10\,ps between the RF and TS clocks,
35~ps in the TOF readout system (front-end and TS-TDC circuits),
10~ps uncertainty in the time-of-flight calculation from $\mu$-track information,
and 13~ps due to the collision position spread of 4.0\,mm because of bunch length.
The 10\,ps uncertainty in the time-of-flight calculation is estimated for $\mu$-pair events (by Monte Carlo),
while it is larger than 10\,ps for hadron tracks due to effects from hadronic
interactions.
The resolution of 110~ps is degraded from the initial resolution of 96\,ps obtained in 1999~\cite{belle-tof}.
Taking into account  the uncertainties discussed above, we extract intrinsic time resolutions of the
Belle TOF detector of 102\,ps and 87\,ps for June of 2008 and 1999, respectively.
The degradation is mainly due to ageing, a reduction of the attenuation length 
and light yield in the TOF scintillation counters over a ten year period.

\section{Beam bunch phase vs bucket number at KEKB}\label{sec:b}
\subsection{Bunch-by-bunch phase detection}\label{subsec:b1}

In the KEKB collider about 450
beam position monitors (BPM) are mounted along the beam pipe for both of the LER and HER, 
in order to provide the beam position at specific locations.
Each BPM consists of four electro-static electrodes to detect beam passages.
One BPM is used to study possible beam phase variation as a function of bucket number in a beam train.
The beam pick-up signal is fed into a gated beam-position monitor system (GBPM) 
that employs an I/Q (In-phase and Quadrature-phase) demodulator operating at the RF frequency of 508.9MHz
to extract two orthogonal signals, $V_{\rm sin}$ and $V_{\rm cos}$ for each bunch. 
A gate module in the front end selects a specific bucket in a beam train and enables 
bunch-by-bunch detection of the beam. 
The two orthogonal components and a specific bucket number
are sampled by a beam revolution clock of approximately 10~$\mu$\,s and stored in memory by GBPM. 
The beam phase $\phi_{\rm beam}$ of a specific bunch, with respect to its bucket phase $\phi_{\rm RF}$ 
synchronous to the 508.9\,MHz RF clock, is calculated  
from the two components $V_{\rm sin}$ and $V_{\rm cos}$ by:
\begin{equation}
\phi_{\rm beam}-\phi_{\rm RF}=\tan^{-1}(-\frac{V_{\rm sin}}{V_{\rm cos}}).
\label{eq3}
\end{equation}
When the GBPM data are averaged over 2,000 turns, we estimate 
a resolution of $\pm0.1$ degrees in the phase measurement
including the jitter of the RF reference clock. This corresponds to a time resolution of about $\pm0.5$ ps.

\subsection{Transient beam loading effect}\label{subsec:b2}
The amplitude and phase of the RF in the accelerating cavities is modulated by the BAG, since the beam loading 
is different between the BAG and the beam train occupied by bunches. As a result, the synchronous position is shifted 
bunch-to-bunch in a beam train. Assuming that the cavity is operated at the optimum tuning and that the filling time of the cavity   
{$T_{f}$} is much longer than the revolution period  $T_{0}$, i.e. $T_0/T_f$ $\ll 1$, the beam-phase difference $\Delta\phi_{\rm HT}$
between the head and tail bunches is given approximately in ref.~\cite{kekb3} as:
\begin{equation}
\Delta\phi_{\rm HT}=\frac{I_t}{2V_c}\frac{R_s}{Q}\omega_{\rm RF}\Delta{t_g},
\label{eq4}
\end{equation}
where $I_t$ is the total beam current, $R_s$ is the shunt impedance,  $V_c$ is the cavity voltage,  $Q$ is the Q-value of the cavity,   
$\omega_{\rm RF}$ is the angular RF frequency and $\Delta t_g$ is a transit time of one BAG. 
In KEKB, the LER operates using ARES cavities with ${T_0}/{T_f}=0.54$, while the HER operates
using two different types of cavities, ARES and SC (about a half of the ARES ${T_0}/{T_f}$ value). 
This causes the LER to have stronger non-linear behavior of the beam phase.

\begin{figure}[!htb]
\centering
\includegraphics[width=0.60\textwidth]{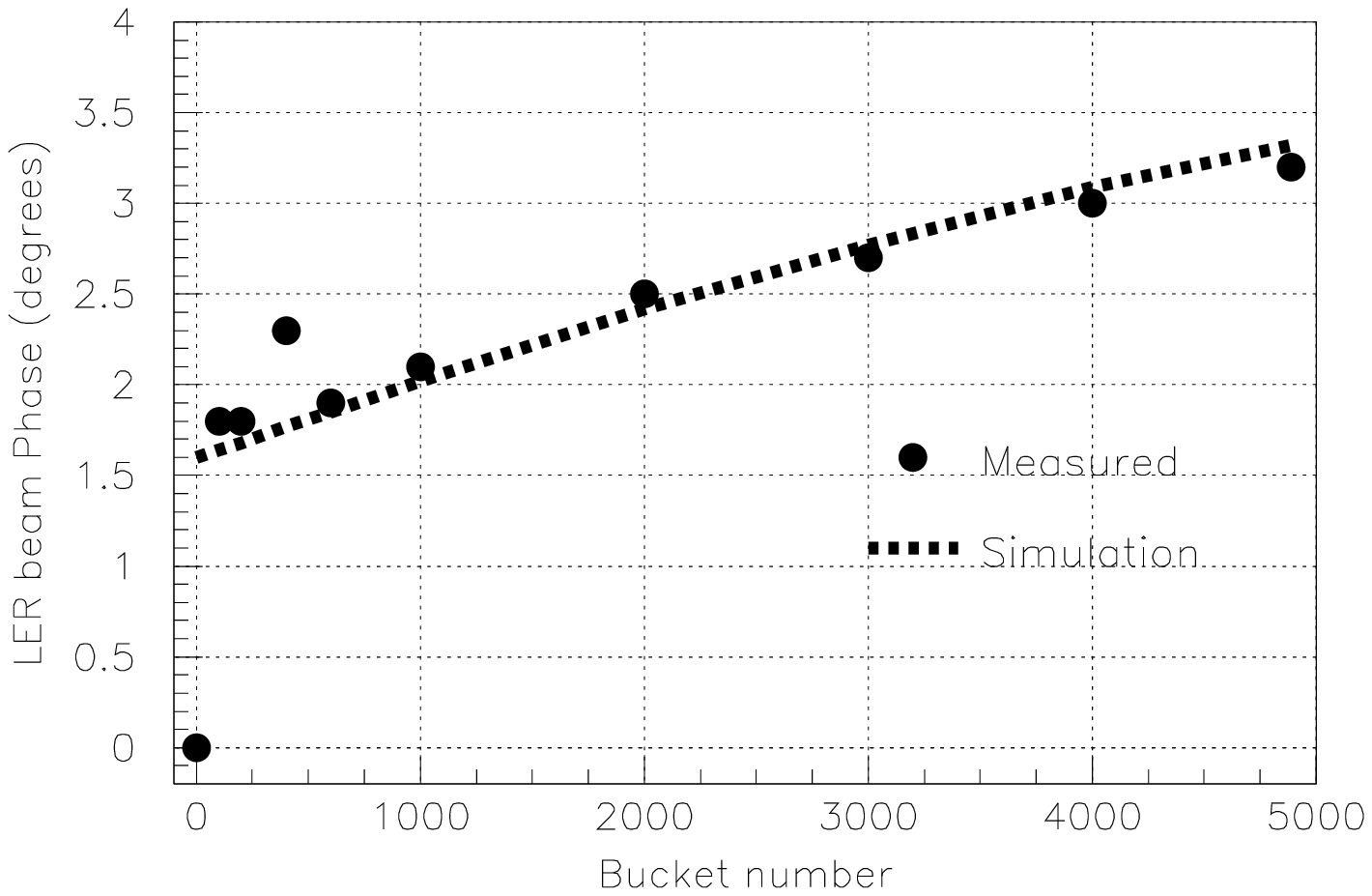}
\centering
\caption{
{Beam phase (in degrees) as a function of bucket number. The solid circles are the measured values
while the dashed line shows the results of simulation.} 
}
\label{fig:beamsim}
\end{figure}

{In order to evaluate the modulation due to the BAG, a time domain  
simulation code was developed, taking into account beam loading  
on the accelerating mode of the cavities~\cite{kekb4}. The simulation result  
was compared with the LER beam phase measurement, as shown in Fig.~\ref{fig:beamsim} 
of Ref. ~\cite{kekb5}.
The simulation (dashed line) accurately reproduces the data (closed circles) except for the first bucket.
The rapid change in the beam head in the data is thought to be due to low Q longitudinal  
wakes that are not taken into account in the simulation.}

\bf{References}


\end{document}